\documentclass{jfm}
\usepackage{graphicx}
\usepackage{enumitem}
\usepackage{epstopdf, epsfig}
\usepackage{amsmath}
\usepackage{mathtools}
\usepackage{cite}
\usepackage{xcolor}
\usepackage[colorlinks=true,
            linkcolor=blue,
            urlcolor=blue,
            citecolor=blue]{hyperref}

\usepackage{svg}
\shorttitle{Dynamics of Sedimenting Mass Polar Spheroids}
\shortauthor{K. Nissanka, X. Ma, and J. C. Burton}

\title{Dynamics of Mass Polar Spheroids During Sedimentation}

\author{Kavinda Nissanka\aff{1}
  \corresp{\email{knissan@emory.edu}},
  Xiaolei Ma\aff{2}
 \and Justin C. Burton\aff{1}}

\affiliation{\aff{1}Department of Physics, Emory University,
Atlanta, GA 30030, USA
\aff{2}Department of Mechanical Engineering and Materials
Science, Yale University, New Haven, CT 06511, USA}

\begin{document}

\maketitle

\begin{abstract}
The dynamics of sedimenting particles under gravity are surprisingly complex due to the presence of effective long-ranged forces. When the particles are polar with a well-defined symmetry axis and non-uniform density, recent theoretical predictions suggest that prolate objects will repel and oblate ones will weakly attract. We tested these predictions using mass polar prolate spheroids, which are composed of 2 mm spheres glued together. We probed different aspect ratios ($\kappa$) and center of mass variations ($\chi$) by combining spheres of different densities. Experiments were done in both quasi-two-dimensional (2D) and three-dimensional (3D) chambers. By optically tracking the motion of single particles, we found that the dynamics were well-described by a reduced mobility matrix model that could be solved analytically. Pairs of particles exhibited an effective repulsion, and their separation roughly scaled as ~$(\kappa - 1)/\chi^{0.39}$, i.e. particles that were more prolate or had smaller mass asymmetry had stronger repulsion effects. In 3D, particles with $\chi>0$ were distributed more uniformly than $\chi=0$ particles, and the degree of uniformity increased with $\kappa$, indicating that the effective 2-body repulsion manifests for a large number of particles.   

\end{abstract}


\section{Introduction}

Sedimentation is a longstanding and important problem in fluid dynamics. In its simplest form, particles far from equilibrium settle in a fluid through some external forcing, typically gravity, at low Reynolds number \citep{stokes1851}. Throughout its storied history, one can observe a microcosm of physics problems that span multiple fields. Starting from basic hydrodynamics, the long range velocity fields generated by sedimenting particles lead to several interesting phenomena \citep{stokes1851,Brady1988,Ramaswamy1,Xue1992,guazzellimorris2011}. Examples include unbounded velocity fluctuations \citep{Caflisch}, chaotic behavior \citep{Brady1988,Janosi1997}, and periodic orbits \citep{Ekiel2005,Jung2006,Chajwa2019,claeys_brady_1993_p1}. Sedimentation is found throughout nature; from silt and sand in a river, to biogenic particles in the ocean \citep{Monroy2019}. Most sedimentation work has been done on uniform particles or particles with simple symmetries. But within nature, most particles are not uniform. They can be rough and polygonal, and they can be made of many different materials, causing their mass to be distribted non-uniformly \citep{Domokos18178}. For example, it has been found that some phytoplankton adjust their center of mass to respond to external environmental flows for better survival in turbulent environments \citep{Sengupta2017}. 


Gravitational sedimentation at low Reynolds number (Stokes flow) is a special case of the Navier-Stokes equation where inertia is negligible. Because of this, Stokes flow is quasistatic and time reversible. For a single spherical particle of radius $R$ and density $\rho_p$ settling in an unbounded fluid of density $\rho_f$ and viscosity $\eta$, balancing the Stokes drag force with gravitational and buoyant forces leads to the following expression for the steady state terminal velocity:
\begin{equation}
    U_T=\dfrac{2}{9}\dfrac{\rho_p-\rho_f}{\eta}g R^2.
    \label{terminal}
\end{equation}
Here $g$ is the gravitational acceleration. The addition of many other particles in the fluid complicates this picture. To leading order, the fluid disturbance at a distance $r$ from a sedimenting sphere with velocity $U_{s}$ and radius $R$ scales as $U_{s}R/r$. In sedimenting suspensions of many particles, these long range hydrodynamic interactions complicate a local description of particle dynamics. Batchelor solved the problem of a diverging mean sedimentation velocity \citep{batchelor1972}, but Caflisch and Luke pointed out that the velocity \emph{fluctuations} were still unbounded as the system size increases \citep{Caflisch}. 

To illustrate the Caflisch-Luke paradox, consider the variance of the sedimentation velocity of a group of $N$ particles contained in a volume of size $L$. The volume fraction $\phi$ of particles is $NV_{p}/L^{3}$, where $V_{p} = \frac{4}{3}\pi R^{3}$ is the volume of a single particle. Within this region, if the particles are randomly and independently distributed, the fluctuation in particle number is simply $\sqrt{N}$. To find the velocity fluctuations, we can balance the total change in the Stokes' drag force over the suspension with the change in gravitational and buoyant forces due to these number fluctuations: $6\pi\eta L\Delta v \approx (\rho_p-\rho_f)V_pg\sqrt{N}$. Solving for $\Delta v$, we arrive at the fractional change in velocity, $\Delta v/v_{o} = L^{1/2}\sqrt{\phi R^{2}/V_{p}}$. This would indicate that the velocity fluctuations depend on the system size, $L$. Simulations agree with these predictions in unbounded fluids \citep{Mucha2004,Ladd1996,Ladd1997,Cunha2002,Koch1993,Koch1994}, while experiments generally observe a limit to the size of the fluctuations \citep{Nicolai1995,Segre1997,Xue1992,HAM1988533}. 

To reconcile this paradox, several different physical mechanisms have been proposed. The long-ranged interactions must be screened out by some large length scale, or by changing the interactions themselves. For example, wall effects at the size of the experimental container \citep{Brenner1999}, correlated particle positions arising from a pre-imposed structure factor  \citep{koch_shaqfeh_1989,koch_shaqfeh_1991}, polydispersity \citep{Nguyen2005}, stochasticity in the concentration \citep{Levine1998}, stratification \citep{Mucha2004}, or shape effects \citep{Goldfriend2017,Witten_2020,Chajwa2019,Chajwa2020,Krapf2009,Doi2005,Palusa2018}. The latter example is of particular interest since it is a local change to particle interactions. Shape effects can be captured within Stokes flow using a response matrix that only depends on particle geometry and couples to external forces and torques. 

We start by considering the Navier-Stokes equation for an incompressible fluid in the low Reynolds number regime:  
\begin{gather}
    \vec{\nabla} P = \eta\vec{\nabla}^2 \vec{\bf v} + \vec{\bf f}_{b}, \\ \space \vec{\nabla}\cdot \vec{\bf v} = 0,
    \label{stokes}
\end{gather}
where $P$ is the pressure, $\eta$ is the dynamic viscosity, $\vec{\bf v}$ is the velocity field, and $\vec{\bf f}_{b}$ are any body forces per unit volume on the fluid, such as gravity. The linearity of these equations allows us to write the equations of motion for a single particle suspended in the fluid and subjected to an external force or torque as:
\begin{gather}
    \vec{v}(t) = \boldsymbol{T}_{vF}\cdot\vec{F} + \boldsymbol{T}_{v\tau}\cdot\vec{\tau},\\ 
    \vec{\omega}(t) = \boldsymbol{T}_{\omega F}\cdot\vec{F} + \boldsymbol{T}_{\omega \tau}\cdot\vec{\tau},
\end{gather}
which can be written in matrix form as:
\begin{equation}
		    \begin{pmatrix}
		    \vec{v} \\
		    \vec{\omega}
		    \end{pmatrix}
		    =
		    \begin{pmatrix}
		    \boldsymbol{T}_{vF} & \boldsymbol{T}_{v\tau}\\
		    \boldsymbol{T}_{\omega F} & \boldsymbol{T}_{\omega\tau}
		    \end{pmatrix}
		    \begin{pmatrix}
		    \vec{F} \\
		    \vec{\tau}
		    \end{pmatrix}.
    \label{MM-gen}
\end{equation}
Here $\vec{\bf \omega}$ is the angular velocity of rotation about the center of geometry, $\vec{F}$ and $\vec{\tau}$ are the external forces and torques, respectively. The convention we use is the same as \citet{Witten_2020}. The shape dependent $\boldsymbol{T}$ matrices couple the velocities of the particle to external forces and torques. In the fixed lab frame, the matrices depend on the particle's orientation to the imposed flow. We can also put restrictions on the matrices by physical insight. The dissipated power of the object, $\vec{F}\cdot\vec{v} + \vec{\tau}\cdot\vec{\omega}$, must be positive, which implies the diagonal blocks, $\boldsymbol{T}_{vF}$ and $\boldsymbol{T}_{\omega\tau}$ must be symmetric, and $\boldsymbol{T}_{v\tau}$ and $\boldsymbol{T}_{\omega F}$ must be transposes of each other but not necessarily positive or symmetric. Taken together, these matrices comprise the \emph{mobility matrix} $\boldsymbol{T}$ of an object. If you invert the relation, the matrix is called the resistance matrix. As an illustration, for a uniform sphere in an unbounded fluid, the mobility matrix is:
\begin{equation}
		    \begin{pmatrix}
		    \vec{v} \\
		    \vec{\omega}
		    \end{pmatrix}
		    =
		    \begin{pmatrix}
		    \frac{1}{6\pi\eta R}\delta_{ij} & 0\\
		    0 & \frac{1}{8\pi\eta R^{3}}\delta_{ij}
		    \end{pmatrix}
		    \begin{pmatrix}
		    \vec{F} \\
		    \vec{\tau}
		    \end{pmatrix},
    \label{MM-sphere}
\end{equation}
where $\delta_{ij}$ is the Kronecker delta. 


The dynamics of a single particle are determined by the time evolution of $\boldsymbol{T}$. As the particle moves through the fluid, its orientation can change with respect to the center of mass velocity. The orientation of the particle relative to the force determines what $\boldsymbol{T}$ looks like in the lab frame. Analogously, if we move to the body frame of the particle, $T$ becomes fixed and the force and torque become time dependent. The motion of the particle cannot change the magnitude of the force, so only the force's direction changes with time. Depending on the symmetries of $\boldsymbol{T}$, different classes of trajectories can be found. For a comprehensive list of these trajectories and symmetries, refer to \citet{Doi2005}, \citet{Krapf2009}, and \citet{Witten_2020}.


In the case of gravitational sedimentation, asymmetric particles with mass distribution polarity will undergo rotation in response to external forcing \citep{Witten_2020}. This is because the total form and skin drag on the particle can apply a net torque when the center of mass is in a different location than the geometric center of the particle. Consequently, an external force leads to a net torque, and the particle will rotate so that the external force is parallel to an eigendirection of $\boldsymbol{T}_{\omega F}$ \citep{Witten_2020}. The response of a single particle can have important implications for the sedimentation dynamics of many particles. Recent work has theoretically explored the sedimentation of ``mass polar" prolate spheroids, whose center of mass lies along the major axis away from the geometric center \citep{Goldfriend2017}. These particles are defined by two parameters: the ratio of major to minor axes, $\kappa$, and the center of mass offset from the geometric center, $\chi$. Using a linear stability analysis of a uniform suspension of particles in Stokes flow, they predicted a repulsive interaction for $\kappa>1$ (prolate), and an attractive interaction for $\kappa<1$ (oblate). The effect is surprisingly enhanced for smaller values of $\chi$. These effects, over a large collection of particles, can either enhance particle clustering and velocity fluctuations ($\kappa<1$), or inhibit them ($\kappa>1$). 


Inspired by \citet{Goldfriend2015,Goldfriend2016,Goldfriend2017}, we experimentally tested these predictions by fabricating prolate, mass polar ``dimers" and ``trimers." The particles were composed of multiple spheres of varying materials bonded together. Our experiments tracked the position and rotation of pairs of particles in a quasi-2D environment. First, we examined the motion of single particles in order to quantify the mobility matrix. Using the symmetry properties of prolate spheroidal particles, we derive an analytic solution for the particle dynamics that shows excellent agreement with the experimental data. Then, by sedimenting pairs of particles in the same quasi-2D environment, we found that prolate particles experienced an effective repulsion that increased with $\kappa$ and decreased with $\chi$, in agreement with \citet{Goldfriend2017}. Finally, we sedimented hundreds of particles in a 3D container and analyzed the distribution of their post-sedimented positions. The inherent repulsion manifested as wider spatial distributions of particles on the floor of the experimental apparatus. This shows local changes in particle interactions have a large effect on global sedimentation patterns.

\section{Experimental Methods and Particle Fabrication}

\label{methods}

\begin{figure}
    \centerline{\includegraphics[width = 3in]{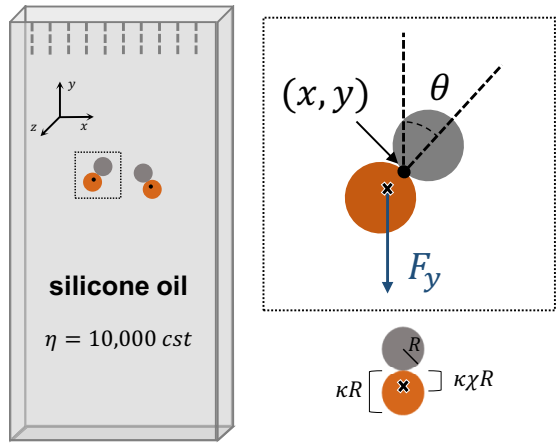}}
    \caption{Schematic diagram of our quasi-2D experimental setup. The tank dimensions are 19 cm $\times$ 15 cm $\times$ 0.4 cm. The top of the tank has a gating mechanism that allowed us to drop multiple particles simultaneously. The mechanism consists of a slotted piece of acrylic and a metal rod in a U shape. By moving the prongs of the rod, the horizontal part can be rotated out of the plane, releasing the particles. simultaneously The schematic on the right depicts a $\kappa=2$ composite particle and coordinates in the lab frame. $\theta$ is defined as the angle between the composite particle's major axis and the vertical direction. The orange sphere has a larger mass density in this case, so the center of mass is shifted away from the center of geometry (black cross) to the position indicated by the black cross. On the bottom right, a physical representation of $\kappa$ and $\chi$ is shown on an example particle. The center of mass of the particle is offset by an amount $\kappa\chi R$. For all our experiments, $R=1$ mm, and the typical Reynolds number was $\sim10^{-4}$.}
    \label{fig:2D Tank}
\end{figure}

Composite particles were fabricated by gluing together smooth ball bearings using a cyanoacrylate based glue . Each sphere had a diameter of 2 mm, and the material and mass density of each sphere were chosen to produce various numerical values of $\chi$. We used the minimal amount of glue possible to adhere the spheres by applying a low-viscosity glue instead of a viscous glue. The remaining thin layer of glue that extended away from the contact point possibly affected the motion of the sedimentation of the particles, but the repeatability of the experiments indicates that this has only a minimal effect. The materials used were aluminum, stainless steel, copper, tungsten carbide, zirconium dioxide, and Delrin. Spheres were glued in either a dimer ($\kappa = 2$) or linear trimer ($\kappa = 3$) configuration. The accessible range of $\chi$ was 0.0-0.43. To analytically calculate $\chi$ for any linear chain of $n$ spherical particles, we assumed all particles were ``light" with density $\rho_l$ except for a single ``heavy" particle with density $\rho_h$ positioned at the end of the chain. The result is: 
\begin{gather}
    \chi = \dfrac{1}{n}\dfrac{(n-1)|\rho_h - \rho_l|}{\rho_h + (n-1)\rho_l }, \hspace{12 pt}n\geq 2.
\end{gather}
The center of mass is displaced by a distance $\kappa\chi R$, for a physical representation of $\kappa$ and $\chi$, see Fig.\ \ref{fig:2D Tank}.

Two sets of experiments used a quasi-2D tank made out of cast acrylic (Fig.\ \ref{fig:2D Tank}). We laser cut sheets of cast acrylic and used SCIGRIP® 4 acrylic plastic cement to glue them together to create a tank of dimensions 19 cm high, 15 cm wide, with a gap of thickness 4 mm. The tank was filled with pure silicone oil of kinematic viscosity 10,000 cSt and density of 0.971 g/cm$^3$. A gating mechanism was placed at the top of the chamber consisting of a thin rectangle of acrylic with 2.5 mm holes spaced out evenly. The holes helped to align the particles so that the initial orientations are fixed before sedimentation. A thin metal rod held them in place and facilitated a simultaneous release of the particles at the beginning of an experimental run.


\begin{table}
    \begin{center}
    \begin{tabular}{ccccc}
        \hline\hline
         Material Combinations & $\rho_l$ (g/cm$^3$) & $\rho_h$ (g/cm$^3$) & $\kappa$ & $\chi$  \\
         \hline
         St+St & $7.82$ & $7.82$ &$2$ & $0$ \\
         \hline
         Cu+St & $7.82$ & $8.92$ & $2$& $0.033$\\
         \hline
         St+ZrO$_{2}$ & $5.68$ & $7.82$ & $2$& $0.080$\\
         \hline
         Cu+ZrO$_{2}$ & $5.68$ & $8.92$ & $2$& $0.11$\\
         \hline
         Al+Pl & $1.42$ & $2.79$ & $2$& $0.16$\\
         \hline
         Tc+St & $7.82$ & $15.63$ & $2$& $0.17$ \\
         \hline
         St+Al & $2.79$ & $7.82$ & $2$& $0.24$\\
         \hline
         Cu+Pl & $1.42$ & $8.92$& $2$& $0.36$\\
         \hline
         Cu+St+St & $7.82$ & $8.92$& $3$& $0.030$\\
         \hline
         Tc+Cu+Cu & $8.92$ & $15.63$ & $3$& $0.11$\\
         \hline
         St+Al+Al & $2.79$ & $7.82$ & $3$& $0.18$\\
         \hline
         Cu+Pl+Pl & $1.42$ & $8.92$& $3$& $0.25$ \\
         \hline
         St+St+St & $7.82$ & $7.82$& $3$& $0$ \\
         \hline
    \end{tabular}
    \end{center}
    \caption{The different types of particles used in out experiments along with their corresponding $\kappa$ and $\chi$ values. Materials used are: steel (St), aluminum (Al), copper (Cu), Delrin plastic (Pl), tungsten carbide (Tc), and zirconium dioxide (ZrO$_{2}$). Values of $\chi$ are kept to two significant digits.}
    \label{tab:particle types}
\end{table}

After the particles were released, we imaged their sedimentation using a CCD camera (Point Grey) at 6 frames per second with a spatial resolution of 12 pixels per mm. After recording, we processed the images using ImageJ \citep{Schindelin2012} for easier detection of each sphere in a composite particle. Images were first binarized with a brightness threshold, then each sphere was separated with a watershedding algorithm. The resulting image was eroded, leaving us with easily-trackable objects composed of white pixels. Particle tracking and linking between frames were done with TrackPy \citep{allan_daniel_b_2021_4682814}. The resulting trajectories of the individual spheres were used to calculate various quantities associated with the dynamics of the composite particles. 

The second set of experiments were done in a cylindrical 3D chamber of diameter of 12 cm and a height of 21 cm (see Sec.\ \ref{3Dsed}). The chamber was fabricated from a cast acrylic tube with wall thickness 12 mm. The chamber was also filled with silicone oil of the same viscosity (10,000 cSt). We placed 100 particles of a single $\kappa$ and $\chi$ combination in the fluid and sealed the chamber so that there were no trapped air bubbles. Particles were allowed to sediment under gravity to the bottom of the chamber, and the distribution of particles was imaged from above. We then flipped the chamber and repeated the experiment 50 times for each set of particles. Due to finite-size wall effects driving convection and particles resting on top of one another, identifying the individual spheres from each particle was not feasible, as done in the 2D experiments. Thus, images were cropped and binarized and the spatial distributions of black pixels were analyzed. 

The quasi-2D geometry allows us to easily track the position and rotation of particles, but it also imposes a form of screening for the interactions between particles. The divergence of velocity fluctuations in suspensions arises from the $1/r$ decay of velocity around a sedimenting particle, however, in confined 2D environments the fluid flow decays as $1/r^2$. A detailed discussion of the differences can be found in \citet{beatus2012}. The faster decay allows convergence of the velocity fluctuations found in 3D, meaning that the majority of the screening is provided by the confining walls of our chamber. Although this is important for a statistically large number of particles, our results show that mass polarity strongly affects sedimentation dynamics in both 2D and 3D geometries.


\section{Results and Discussion}

\subsection{Single Particle Dynamics}

\begin{figure}
    \centerline{\includegraphics[width = 5in]{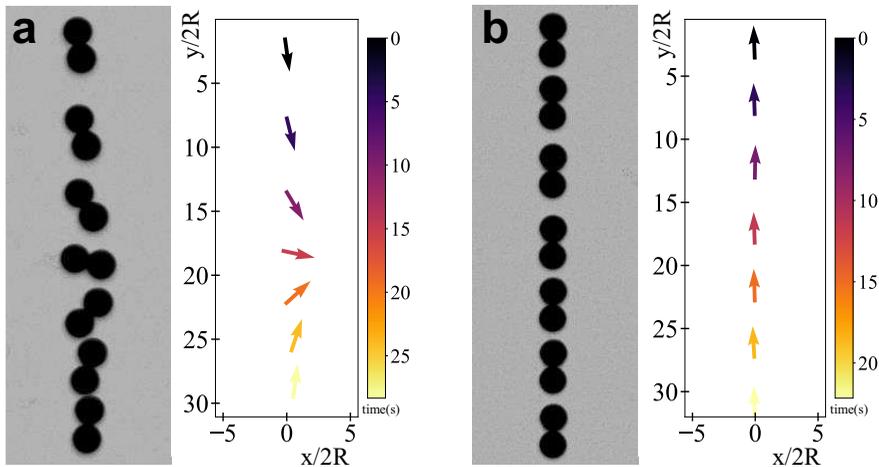}}
    \caption{Two representative examples of the particle trajectories in our single particle experiments. $x=0$ is defined as the geometric center of the particle at the earliest time. Panel (a) shows a particle with $\chi > 0$ (Cu+St, see Table \ref{tab:particle types}). The left part of the panel is a composite image of the particle during the length of the experiment. The right graph shows the corresponding particle orientations, with the arrows pointing from the heavier sphere (Cu) to the lighter sphere (St). The color bar represents time. Gravity points downward in all pictures. Panel (b) shows a particle with $\chi = 0$  (St+St).}
    \label{fig:single particle time trace}
\end{figure}

After fabricating the composite, prolate particles, we observed the sedimentation of single, isolated particles to better understand their dynamics and to extract the terms in the mobility matrix (Eq.\ \ref{MM-gen}). The response of a single particle to an external force or torque informs its effective interactions with neighboring particles \citep{Witten_2020,Goldfriend2017}.  For example, a rod-shaped particle of uniform mass density will sediment without a change in its initial angle \citep{Ramaswamy1, Witten_2020}. This results in a diagonal drift. However, the mass polarity of our objects causes them to align with the external gravitational field, meaning that a mass polar object will rotate until its center of geometry lies directly above its center of mass ($\theta = 0$). For our experiments, mass polar particles were released from an initial angle of $\theta = \pi$, so that they rotated a total of $\pi$ radians throughout the sedimentation process. A trajectory for a single $\kappa$ = 2 particle composed of Cu+St (see Table \ref{tab:particle types}) is shown in Fig.\ \ref{fig:single particle time trace}a. Particles with larger values of $\chi$ rotated much more rapidly due to the larger gravitational torque applied to the geometric center of the particle. This can be compared with a St-St particle in Fig.\ \ref{fig:single particle time trace}b, which shows no preference for rotations since it has no mass polarity ($\chi=0$). For particles with $\chi=0$, we occasionally observed ``fluttering", or oscillations of angular orientation during sedimentation. This was likely due to interactions with the walls of the experimental chamber during slight rotations out of the quasi-2D plane of the experiment \citep{Mitchell2014,D'Angelo2017}.


To quantitatively capture the coupling between the external force and dynamics of single particles, we applied the mobility matrix formalism (Eq.\ \ref{MM-gen}). Because we are using a quasi-2D geometry, the complexity of the problem is reduced since the particle can only rotate in the plane. However, the mobility coefficients will be different from those measured in an unbounded, 3D fluid. With two planar walls, our experimental setup is most similar to a Hele-Shaw cell, where the mobility matrix formalism has already been successfully implemented \citep{Bet_2018} and tested \citep{Georgiev21865}.
Because we are considering symmetric prolate particles, the mobility matrix in the body frame (indicated by superscript $b$) is reduced to:
\begin{equation}
    \begin{pmatrix}
    v_{x}^b\\
    v_{y}^b\\
    \omega_{z}^b
    \end{pmatrix}
    =
    \dfrac{1}{6\pi\eta R}
    \begin{pmatrix}
    a_t & 0 & 0\\
    0 & b_t & 0\\
    0 & 0 & \dfrac{3a_r}{4R^2}
    \end{pmatrix}
    \begin{pmatrix}
    F_{x}^b\\
    F_{y}^b\\
    \tau_{z}^b
    \end{pmatrix},
    \label{MM-spec}
\end{equation}
where $v_{x}^b$ and $v_{y}^b$ are the translational velocities in the body frame and $\omega_{z}^b$ is the angular velocity perpendicular to the plane of motion. $F_{x}^b$ and $F_y^b$ are the components of the gravitational force in the body frame, and $\tau_{z}^b$ is the external torque from gravity about the particle's center of geometry (see Fig.\ \ref{fig:2D Tank}). The dimensionless translational mobility coefficients $b_t$ and $a_t$ represent mobility along the major and minor axes of the particle ($b_t>a_t$). The dimensionless rotational mobility coefficient is $a_r$. These coefficients should be identical for all of our particles with the same $\kappa$ and $R$, regardless of the internal density distribution ($\chi$). They characterize the drag from the external flow, which applies stress on the surface of the particle.

Our experimental data, however, are collected in the lab frame. Thus, we first rotate all vectors and the mobility matrix by an angle $\theta$ (Fig.\ \ref{fig:2D Tank}) to obtain the equations of motion in the lab frame:
\begin{gather}
    \boldsymbol{\Omega} = 
    \begin{pmatrix}
    \cos(\theta) & \sin(\theta)& 0\\
    -\sin(\theta) & \cos(\theta) & 0\\
    0 & 0 & 1
    \end{pmatrix}
\end{gather}
\begin{gather}
    \boldsymbol{\Omega}\cdot
    \begin{pmatrix}
    v_{x}^b\\
    v_{y}^b\\
    \omega_{z}^b
    \end{pmatrix} = 
    \left(\boldsymbol{\Omega}\cdot
    \dfrac{1}{6\pi\eta R}
    \begin{pmatrix}
    a_t & 0 & 0\\
    0 & b_t & 0\\
    0 & 0 & \dfrac{3a_r}{4R^2}
    \end{pmatrix}
    \cdot\boldsymbol{\Omega}^{-1}\right) \boldsymbol{\Omega}\cdot
    \begin{pmatrix}
    F_{x}^b\\
    F_{y}^b\\
    \tau_{z}^b
    \end{pmatrix}
\end{gather}
After multiplying and collecting terms, we use the substitutions $2c_1 = a_t + b_t$, $2c_2 = b_t - a_t$, and $c_3 = 3a_r/4$ to write the result in the following form:
\begin{gather}
    \begin{pmatrix}
    v_{x}\\
    v_{y}\\
    \omega_{z}
    \end{pmatrix}
    =\dfrac{1}{6\pi\eta R}
    \begin{pmatrix}
    c_{1} - c_{2}\cos(2\theta) & c_{2}\sin(2\theta) & 0\\
    c_{2}\sin(2\theta) & c_{1} + c_{2}\cos(2\theta) & 0\\
    0 & 0 & \dfrac{c_{3}}{R^2}
    \end{pmatrix}
    \begin{pmatrix}
    0\\
    F_{y}\\
    \tau_{z}
    \end{pmatrix}.
    \label{MM-spec-expand}
\end{gather}
We have chosen this parameterization out of convenience. For example, in the case of a perfect sphere, $b_t=a_t$, thus $c_1=1$, $c_2=0$, and $c_3=3/4$ (Eq.\ \ref{MM-sphere}). We have dropped the superscript since we are referring to the lab frame where the gravitational force only points in the $y$-direction.
The matrix multiplication above gives us the following equations of motion for our particles in the lab frame:
\begin{gather}
    v_{x}=\dot{x} = \frac{c_{2}\sin(2\theta)}{6\pi\eta R}F_{y}, \label{MM-full1}\\
    v_{y}=\dot{y} = \frac{c_{1} + c_{2}\cos(2\theta)}{6\pi\eta R}F_{y}, \label{MM-full2}\\
    \omega_{z}=\dot{\theta} = \frac{c_{3}}{6\pi\eta R^{3}}\tau_{z}.
    \label{MM-full3}
\end{gather}
\begin{figure}
    \centerline{\includegraphics[width = 4in]{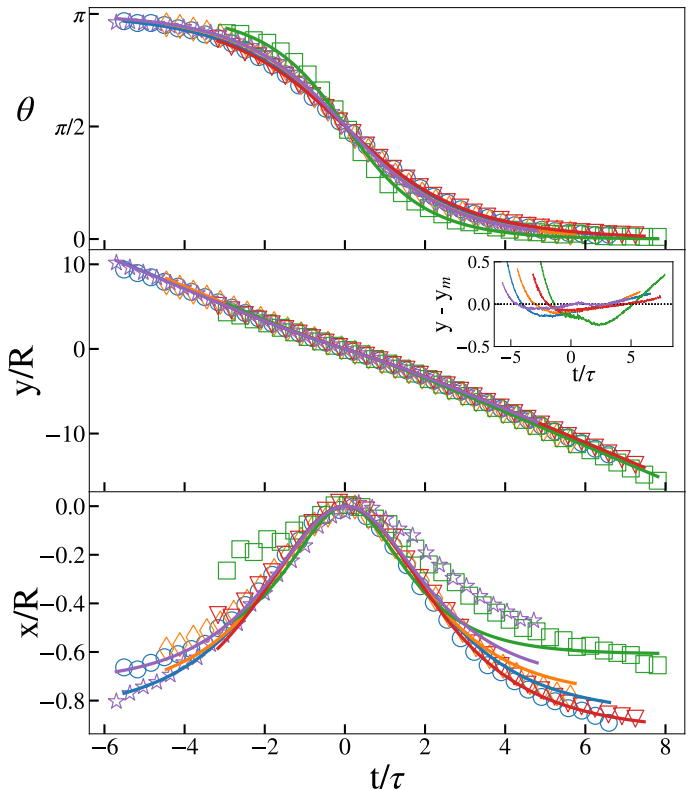}}
    \caption{Data for 5 experiments with a single Al+Pl particle (Table \ref{tab:particle types}). Only 5\% of points are plotted for clarity. Initially, the heavy aluminum sphere begins above the lighter Delrin sphere. Open symbols represent data, and curves are model fits from Eqs.\ \ref{theta-t}, \ref{y-t}, and \ref{x-t}. Different symbols and colors are separate experiments. Inset: residual difference between the model fit $y_m$ and the data $y$ for the vertical position of the particle.}
    \label{fig:single_particle}
\end{figure}
\begin{figure}
    \centering
    \centerline{\includegraphics[width=\columnwidth]{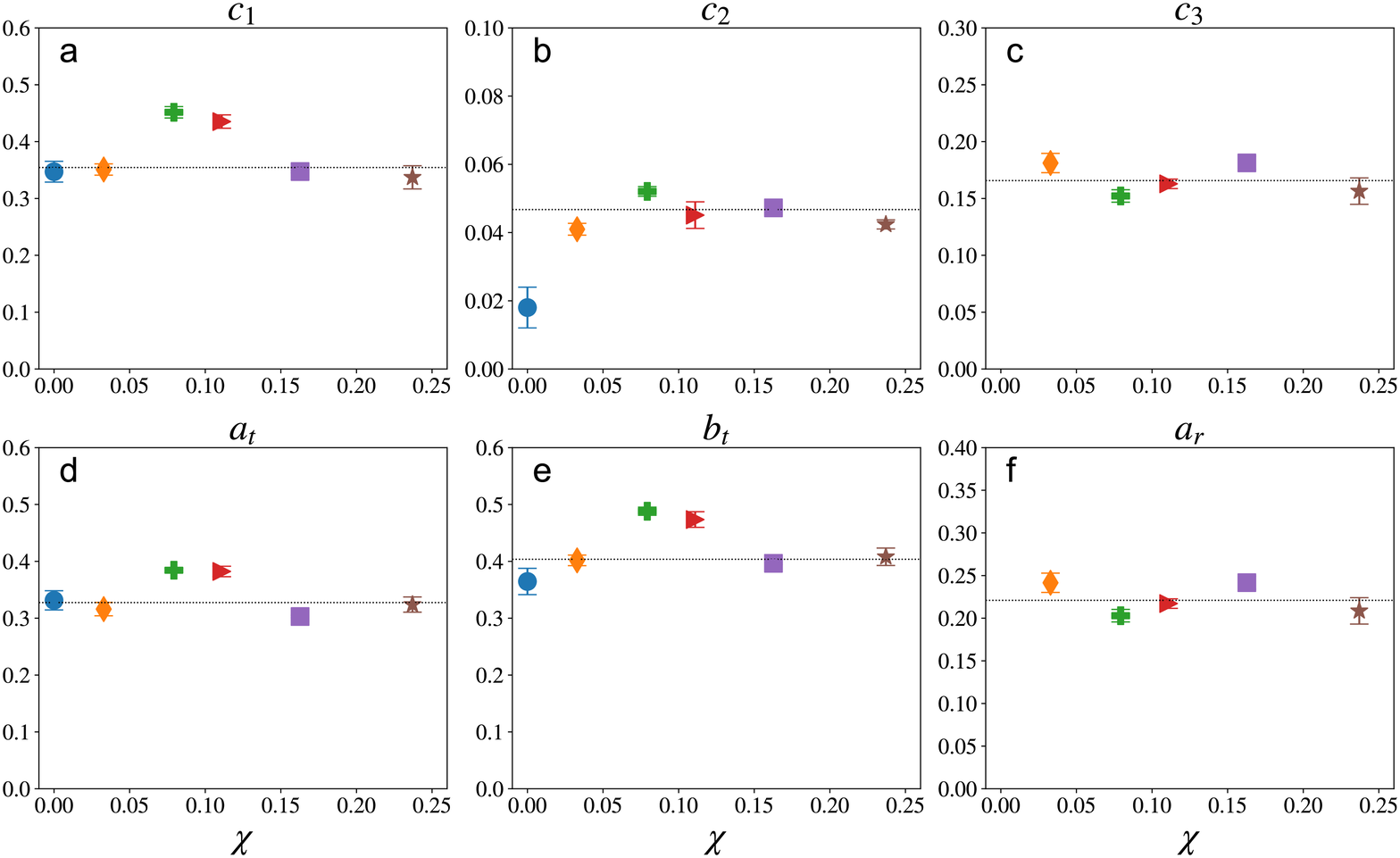}}
    \caption{Best-fit parameters vs. $\chi$ from Eqs.\ \ref{theta-t}-\ref{y-t} for single particle sedimentation experiments with $\kappa=2$. The material combinations used were: circle, St+St; diamond, Cu+St; plus, St+ZrO$_{2}$; right triangle, Cu+ZrO$_{2}$; square, Al+Pl; star, Al+St (see Table \ref{tab:particle types}). Each data point is the weighted mean of five different trials with error bars representing the standard error of the mean. Panels (a-c) show the parameters $c_1$-$c_3$ directly computed from the nonlinear regression of data in the lab frame. Panels (d-f) show the body frame coefficients: $a_t$, $b_t$, and $a_r$. St+St is missing from $c_3$ and $a_r$ because of the limiting form of $\theta(t)$ when $\chi$ = 0 (Eq.\ \ref{theta-ts}), which has no dependence on $c_3$.}
    \label{fig:single particle params}
\end{figure}

The dotted variables denote differentiation with respect to time. Similar simplified equations for single particle dynamics in quasi-2D geometries have been derived by \citet{Bet_2018} and \citet{Ekiel_Je_ewska_2009}. In our experiments, the net force and torque on a particle will depend on the values of $\kappa$ and $\chi$. For $\kappa=2$ particles, the net gravitational force, and torque about the center of geometry are:
\begin{gather}
   F_{y} = -\frac{4}{3}\pi R^{3}(\rho_{h} + \rho_{l} - 2\rho_{f})g \label{force_fy}\\
    \tau_{z} = -\frac{4}{3}\pi R^{4}(\rho_{h} - \rho_{l})g\sin{\theta}. \label{torque_tauz}
\end{gather}
Equations \ref{MM-full1}-\ref{MM-full3} are coupled through $\theta$, and can be solved analytically. However, the solution can be generalized by making the equations dimensionless. We used the sphere radius $R$ for a characteristic length scale, and $\tau=R/U_T$ for the characteristic time scale, where $U_T$ is the terminal velocity of the lighter sphere (Eq.\ \ref{terminal}). This nondimensionalization results in the following equations of motion, where all variables are considered dimensionless for clarity of notation: 
\begin{gather}
    \dot{x} = -K_1c_{2}\sin(2\theta)\label{xnondim}\\
    \dot{y} = -K_1(c_{1} + c_{2}\cos(2\theta))\label{ynondim}\\
    \dot{\theta} = -K_{2}c_{3}\sin(\theta)\label{thetanondim}\\
    K_1 = \frac{\rho_{h} + \rho_{l} - 2\rho_{f}}{\rho_{l} - \rho_{f}}\\
    K_{2} = \dfrac{\rho_{h} - \rho_{l}}{\rho_{l} - \rho_{f}}
    \label{MM-scaled}
\end{gather}
Equation \ref{thetanondim} can be immediately solved since it is independent of the other equations. The result is:
\begin{gather}
    \cot\left(\frac{\theta(t)}{2}\right)= \cot\left(\frac{\theta_{0}}{2}\right)e^{K_2c_{3}t},    
    \label{theta-t}
\end{gather}
where $\theta_0$ is the initial value of $\theta$ at $t=0$. Plugging this back into Eqs.\ \ref{xnondim} and \ref{ynondim} and simplifying algebraically, we get:
\begin{gather}
    x(t) = x_0 + \dfrac{4c_{2}FK_1\cot(\theta_0/2)}{c_{3}K_{2} - c_{3}F^2 K_{2} \cot(\theta_0/2)^2}  -\dfrac{2c_{2}K_1\sin\left(\theta_0\right)}{c_{3}K_{2}},
    \label{x-t}
\end{gather}
\begin{gather}
    y(t) = y_0 - \dfrac{K_1}{K_2 c_3}\left((c_1 + c_2)c_3K_2t + 2c_2\left(\cos(\theta_0)+\dfrac{1-F^2\cot^2\left(\frac{\theta_0}{2}\right)}{1+F^2\cot^2\left(\frac{\theta_0}{2}\right)} \right)\right),
    \label{y-t}
\end{gather}
where $F = e^{K_{2}c_{3}t}$ is a function of time, and used here for compactness. In the limit of particles with uniform mass density ($K_2\rightarrow 0$, $\chi\rightarrow 0$), these functional forms simplify to:
\begin{gather}
    \theta(t)=\theta_0,    
    \label{theta-ts}
\end{gather}
\begin{gather}
    x(t) = x_0 - c_2 K_1 t \sin(2\theta_0),
    \label{x-ts}
\end{gather}
\begin{gather}
    y(t) = y_0 - K_1 t (c_1+c_2\cos(2\theta_0)).
    \label{y-ts}
\end{gather}
Equations \ref{theta-ts}-\ref{x-ts} verify the prediction that for polar particles of uniform density, the angle of inclination doesn't change, and the particle drifts laterally in the $x$-direction \citep{Ramaswamy1}.

After taking the inverse cotangent of Eq.\ \ref{theta-t} and using standard least-squares nonlinear regression, we can fit these analytic forms to the experimental data with very good agreement. Figure \ref{fig:single_particle} shows 5 identical experiments and their corresponding fits. For $\theta(t)$, there are only 2 fitting parameters, $c_3$ and $\theta_0$. Once they are determined by the fit, then $x(t)$ can be fit for the parameters $c_2$ and $x_0$. Finally, $y(t)$ can then be fit for $c_1$ and $y_0$. The curves are compared to each other by assigning $t=0$ when the particles are completely horizontal, i.e. $\theta=\pi/2$. We also moved the $x$ and $y$ origin to correspond to $t=0$. Open symbols represent data, and curves are the fits to Eqs.\ \ref{theta-t}, \ref{y-t}, and \ref{x-t}. The fits for the $x$-position show more systematic deviation from the data, yet the overall displacement is also much smaller. For example, as shown in the inset in $y$ vs. $t$, the residuals of these fits are comparable to the variability in $x$ vs. $t$, which is a fraction of a particle radius in displacement. Although the source of the systematic asymmetry is unclear, we suspect that when particles are released from the gating mechanism, they are not perfectly parallel with the walls of the quasi-2D chamber. If a particle's alignment varies during the rotation from $\theta = \pi$ to $\theta = 0$, we would expect variations in the mobility coefficients (i.e., $c_2$) due to wall effects \citep{Brenner1999,Mitchell2014}, resulting in an asymmetry in $x(t)$ about $\theta = \pi/2$. Additionally, we do not expect errors in particle tracking to lead to systematic asymmetry even though the $x$-motion is on the order of the particle size. Tracking errors would manifest more as random noise rather than systematic deviations from theory.
The data for $\theta$, $x$, and $y$ can also be fit simultaneously using a global least squares regression for all parameters, since parameters appear in multiple equations. We found less than 5\% difference in the fitted parameter values using this method, so we have only chosen to report the results of the sequential fitting. Similar analytic solutions and quality of fits were recently found in the alignment of mirror-symmetric particles in a microfluidic device \citep{Georgiev21865,Bet_2018}. 


One of the major assumptions of our model was that all coefficients are independent of $\chi$, and only depend on the shape of the composite, prolate particles. This is evident from Eq.\ \ref{MM-spec}, since $a_t$, $b_t$, and $a_r$ are dimensionless coefficients that only depend on the particle shape, not the density distribution. We confirmed this prediction using all fits of single particle experiments with $\kappa$ = 2, as shown in Fig.\ \ref{fig:single particle params}a-c. The coefficients $c_1$, $c_2$, and $c_3$ are computed directly from nonlinear least-squares regression of the data (Eqs.\ \ref{xnondim}-\ref{thetanondim}). For particles with $\chi$ = 0 (uniform density), we used Eqs.\ \ref{theta-ts}-\ref{x-ts} to fit the data. In this form, there is no torque from gravity, so $c_3$ cannot be determined and is not shown. However, $c_1$ and $c_2$ can be determined, but are not very reliable because of experimental artifacts that affect the angle (and thus translational velocity) during sedimentation. These artifacts include small differences in the distribution of glue used between the particles, rotations out of the quasi-2D plane, and other 2D confinement effects such as ``fluttering'' \citep{Mitchell2014,D'Angelo2017,Brenner1999}. For finite $\chi$, the particles rotate significantly due to gravitational torque, and $c_1$, $c_2$, and $c_3$ can be determined reliably. There appears to be some small systematic trend in $c_1$, but the overall variation is small and the data for all parameters is consistent with a constant value over the range $0<\chi<0.25$.

Using $2c_1 = a_t + b_t$, $2c_2 = b_t - a_t$, and $c_3 = 3a_r/4$, we computed the shape-dependent drag coefficients of our symmetric particles in the body frame, as shown in Fig.\ \ref{fig:single particle params}d-e. Here again, $a_r$ cannot be determined from $\chi$ = 0 data, and data with finite $\chi$ are most reliable. The average values of these mobility coefficients with $\chi>0$ are shown by the dashed lines:$\bar{a}_t=0.328\pm0.018$, $\bar{b}_t=0.404\pm0.012$, and $\bar{a}_r=0.221\pm0.008$. We suggest that these experimental values for the mobility coefficients can be compared directly to simulations of particles composed of spheres \citep{garcia2002hydrodynamic}. In comparison to sedimentation in an unbounded, 3D fluid, we expect our measured values of $a_t$, $b_t$, and $a_r$ to be somewhat smaller since the particles experience a larger drag due to the confining walls.

\subsection{Sedimenting Particle Pairs}
\label{sec:sediment_particle_pairs}

\citet{Goldfriend2017} theoretically examined a sedimenting suspension of mass polar spheroids using a continuum linear stability analysis. To briefly summarize their results, they considered a suspension of particles settling due to an external body force $F$ in the direction of gravity in a fluid of viscosity $\eta$. A sinusoidal concentration perturbation was applied in a direction perpendicular to the force with amplitude $c(x)$ and a characteristic wavelength $\lambda$. These fluctuations in the concentration create velocity fluctuations, $U(x)$. Balancing the change in gravitational force of the suspension versus the change in the drag force gives: $c\lambda F \approx U \eta /\lambda$. By solving for the amplitude $U$, we see that $U \sim c\lambda^2 F/\eta$. The indefinite scaling of $U$ with $\lambda$ is a demonstration of the Caflisch-Luke paradox described in the introduction. These slabs of particles will also experience vorticity of the magnitude $U/\lambda \sim c\lambda F/\eta$. For uniform spheres, this will cause a rotation of the sphere, but no drift. However, self-aligning objects will be tilted away from their preferred alignment. This causes a drift in the $x$-direction with velocity $\sim \gamma R c F/\eta$, where $\gamma=\gamma(\kappa,\chi)$ is a proportionality constant determined by the shape and mass distribution of an individual particle. For positive $\gamma$, which requires $\kappa>1$ \citep{Goldfriend2017}, the relative velocity of the particles is positive, meaning they drift away from each other. This is the screening mechanism that stabilizes the suspension. For negative $\gamma$, which requires $\kappa<1$, they drift towards each other, leading to unbounded growth of the instability.

\begin{figure}
    \centerline{\includegraphics[width = \textwidth]{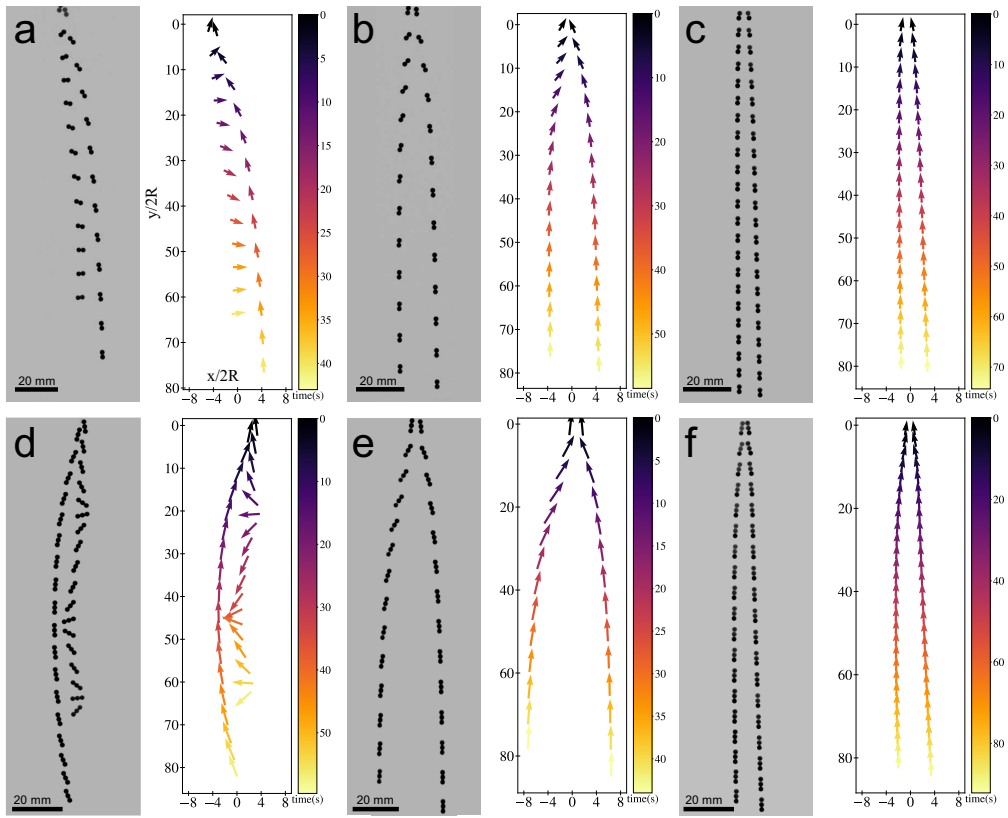}}
    \caption{Experimental trajectories of two-particle interaction experiments. In each alphabetic panel, the image shows a composite of the particles' trajectories during sedimentation. The graph shows the orientation of the particles with arrows pointing from the heavier particle to the lighter particle(s). The color is used to show when two arrows are at the same time during their transit.  $x=0$ is defined as the halfway point between the particle centers on the first frame.  Panels (a), (b), and (c) are $\kappa=2$ particles, panels (d), (e), and (f) are $\kappa=3$ particles. Panels (a) and (d) show particles with $\chi=0$. Panels (b) and (e) show particles with the smallest $\chi$, Cu+St (see Table \ref{tab:particle types}). Panels (c) and (f) show particles with the largest $\chi$, Cu+Pl (see Table \ref{tab:particle types}).}
    \label{fig:2 Particle Exp}
\end{figure}

In our experiments, we examined the particle-level interactions by measuring the relative separation of pairs of prolate ($\kappa>1$) particles as they repel each other during sedimentation.
%
%
%
%
We placed two particles heavy-side down in adjacent slots of the plastic gate so that their initial separation was 3.3 mm. Each experiment was conducted five times for reproducibility. Figure \ref{fig:2 Particle Exp} shows a representative selection of settling trajectories for various values of $\kappa$ and $\chi$. These are composite images of the particles during sedimentation, spaced 3.33 s apart. The arrows to the right of each panel show the orientations of each particle during sedimentation, and the color represents time. First, particles with $\chi = 0$ heavily influenced each other. Their dynamics were typically characterized by one of the particles rotating or flipping completely. This particle often lagged behind the other one, which did not flip, but followed a curved trajectory. This can be seen in both Fig.\ \ref{fig:2 Particle Exp}a and Fig.\ \ref{fig:2 Particle Exp}d. The particles did not preferentially align to gravity, and instead produced a variety of dynamics. For example, the periodic variation in separation visible in \ref{fig:2 Particle Exp}d is reminiscent of Kepler orbits observed in sedimenting pairs of disks \citep{Chajwa2019}. In fact, a periodic variation in the relative position between adjacent, sedimenting prolate particles was theoretically predicted by \cite{claeys_brady_1993_p1} (Fig.\ 4). On average, we did not observe a net repulsion or attraction between our particles with a uniform mass distribution ($\chi=0$).  

For particles with $\chi > 0$, there was an immediate rotation and repulsion between the particles leading to a horizontal separation that grew with time. Eventually the particles would align with the external gravitational field, and the separation saturated. This is shown in Figs. \ref{fig:2 Particle Exp}b-c for $\kappa$ = 2 and Figs. \ref{fig:2 Particle Exp}e-f for $\kappa$ = 3. The finite width of the quasi-2D chamber, 4$R$, introduced a length scale that could potentially set an upper limit on the range of the repulsive interaction. However, we observed that the final separation between the particles could be as much as 30$R$ (Fig.\ \ref{fig:2 Particle Exp}e) for smaller values of $\chi$. 
The repulsive effect was most prominent for particles composed of materials with closely-matched densities (i.e. copper and steel). Although this may seem counter-intuitive at first, particles with $0 < \chi \ll 1$ can rotate away from vertical more easily, and thus experience a larger repulsion and horizontal drift. As $\chi\rightarrow 0$, we expect one of the particles to be able to flip entirely if they are close enough to interact strongly, leading to the periodic type of interactions observed for $\chi=0$ (Figs.\ \ref{fig:2 Particle Exp}a and \ref{fig:2 Particle Exp}d). In this limit, the eventual behavior of the sedimenting particles should be determined both by $\chi$ and by the initial separation.



\begin{figure}
    \centering
    \centerline{\includegraphics[width=3.5in]{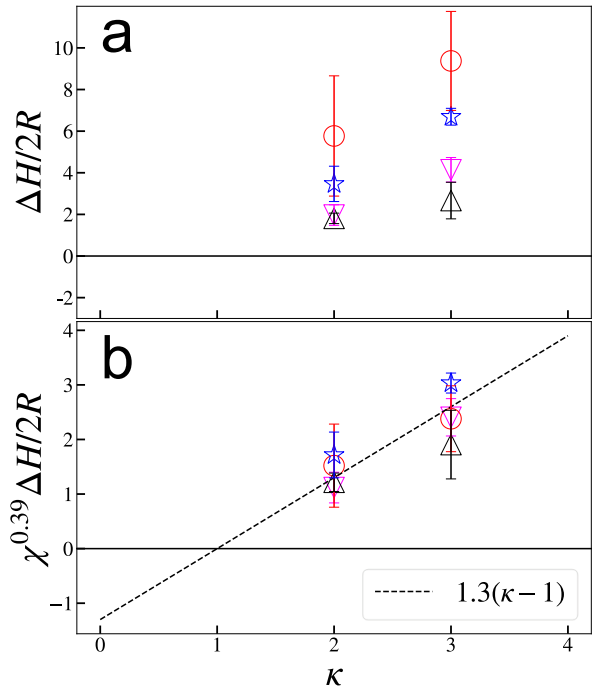}}
    \caption{Graphs of our experimental response parameter, $\Delta H$, versus $\kappa$. $\Delta H$ is defined as the difference between the final and initial horizontal separation. The colors and shapes represent different material combinations of the composite particles. By order of increasing $\chi$, they are Cu+Pl (upright triangle), Al+St (upside down triangle), Tc+St (star), and Cu+St (circle) (see Table \ref{tab:particle types}). Panel (a) is the raw data, with each data point being an average over five runs and error bars representing one standard deviation. Panel (b) is the same data collapsed using the best fit parameters obtained from Eq.\ \ref{sepeq}. }
    \label{fig:response vs. kappa}
\end{figure}

The inverse relationship between $\chi$ and the mutual repulsion was also predicted by \citet{Goldfriend2017}. The authors found that the growth rate of the horizontal velocity fluctuations scaled as $\gamma = \kappa^{2/3}/3\chi$ for highly prolate particles ($\kappa \gg 1$). To quantify this effect in our experiments, we chose to measure the total change in horizontal separation, $\Delta H$, between the particles' geometric centers in each experiment. This is plotted in Fig.\ \ref{fig:response vs. kappa}a as a function of $\kappa$. Generally, the separation increased with $\kappa$. However, in order to compare between  each set of experiments that corresponded to different values of $\chi$, we multiplied the final separation by $\chi^\alpha$, where $\alpha$ was determined by simultaneous fitting of all the data to the following form:
\begin{equation}
\chi^\alpha\dfrac{\Delta H}{2R}=A(\kappa-1),
\label{sepeq}
\end{equation}
where we have imposed the requirement that there be no repulsion for $\kappa=1$ (i.e. single spheres). The fit was performed by subtracting the left and right hand sides of Eq.\ \ref{sepeq}, squaring the difference, and summing over all data points. The best fit values for the parameters were $\alpha=0.39\pm0.05$ and $A=1.28\pm0.20$, where the errors represent one standard error. The fit shows very good agreement with the data, as plotted in Fig.\ \ref{fig:response vs. kappa}b. 

In general, the predictions from \citet{Goldfriend2017} are in excellent qualitative agreement with our experiments, yet the scaling, $\alpha\sim 0.39$, is quite different than that predicted by \citet{Goldfriend2017}: $\alpha\sim 1$. There are a few reasons that can explain this discrepancy: 1) $\gamma$ represents an instantaneous response for an initially-uniform concentration of particles. Here we are using the final separation, $\Delta H$, which is essentially an integral of the repulsion between the particles in time. 2) The quasi-2D environment should screen the long-ranged, $1/r$ hydrodynamic interactions \citep{beatus2012}, so one may expect a different theoretical scaling between $\gamma$ and $\chi$ based purely on geometry. 3) Our quasi-2D chamber may introduce other effects that depend on the thickness of the chamber, for example, it is well known that the net viscous drag force on a sedimenting particle can be dependent on distance to a nearby wall \citep{Brenner1999,Mitchell2014}. 4) Our particles are not perfect examples of the prolate and oblate ellipsoids discussed in \citet{Goldfriend2017}. Despite these differences, the experimental data with different values of $\kappa$ and $\chi$ can be reasonably collapsed using the dependence listed in Eq.\ \ref{sepeq}.

\begin{figure}
    \centering
    \includegraphics[width=5in]{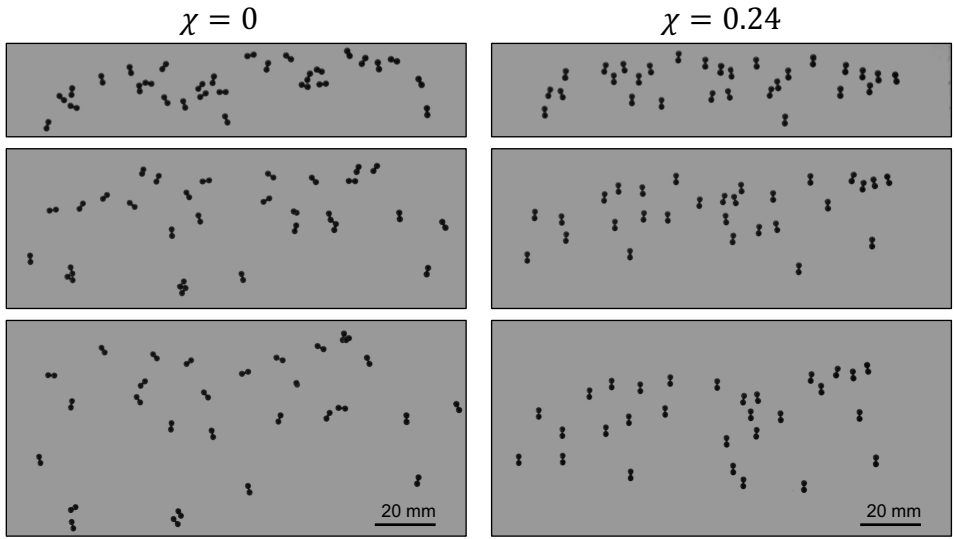}
    \caption{Series of 3 sequential images of 29 particles sedimenting in our quasi-2D chamber. The left column ($\chi$ = 0) shows St+St particles, and the right column ($\chi$ = 0.24) shows St+Al particles (Table \ref{tab:particle types}). The top row shows images at the same vertical position near the top of the experimental chamber at early times, the middle row shows the same particles later in time, and the bottom row shows the particles near the end of the experiment, at the bottom of the chamber.}
    \label{fig:multiparticle 2D}
\end{figure}

Lastly, we verified that this mutual repulsion led to more uniformly distributed suspensions of many particles. We filled our quasi-2D chamber with 29 particles with $\kappa$ = 2. The left column Fig.\ \ref{fig:multiparticle 2D} shows that for particles with $\chi=0$, there is no preferential alignment to gravity, resulting in a large spread of particle separations, both vertically and horizontally. Particle can flip very easily, and often came into contact. Some of the particles experienced small rotations out of the plane as well. The right column of Fig.\ \ref{fig:multiparticle 2D} illustrates that particles with $\chi=0.24$ followed a more uniform spatial distribution. All particles tended to align with gravity, resulting in a mutual repulsion. When particles are in close proximity, they tilted away from the vertical and drifted apart, similar to Fig.\ \ref{fig:2 Particle Exp}. Surprisingly, the particles with $\chi$ = 0.24 did not spread as much in the vertical direction as $\chi$ = 0, suggesting that vertical fluctuations in concentration may be suppressed for $\chi>0$. An intuitive explanation for this behavior stems from the variations in vertical velocities of particles. For $\chi$ = 0, particle rotations lead to a spread in vertical terminal velocities (Eq.\ \ref{y-t}), whereas particles with $\chi>0$ are mostly aligned to gravity and sediment at the same rate. Although vertical fluctuations were not directly addressed in \citet{Goldfriend2017}, we hypothesize that the mutual repulsion in mass polar particles also suppresses the ``clumping instability" observed in uniform suspensions \citep{Chajwa2020} that leads to large vertical separations between particles. 



\section{3D Particle Suspensions}
\label{3Dsed}

\begin{figure}
    \centerline{\includegraphics[width = 4in]{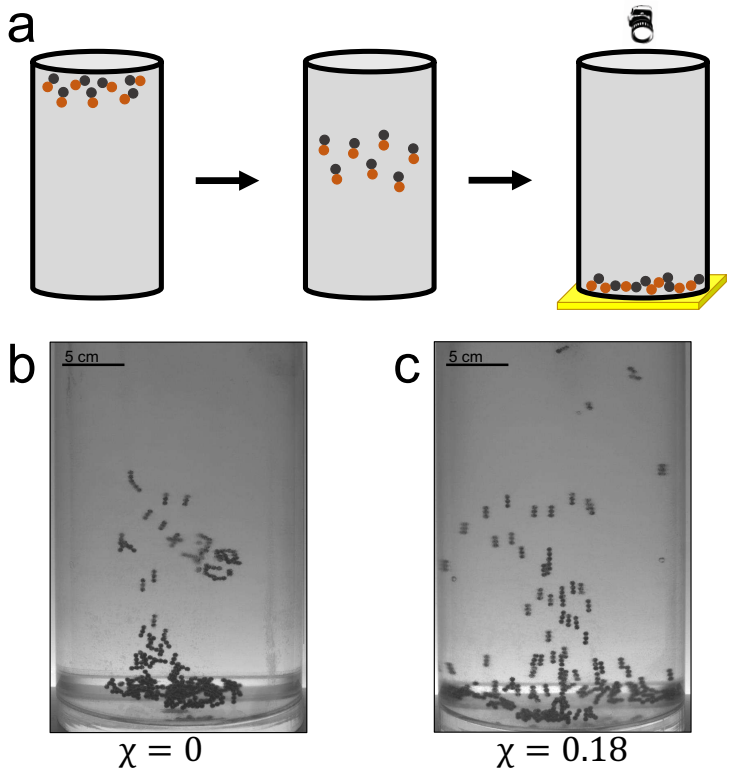} }
    \caption{(a) Experimental procedure for 3D sedimentation. The sealed chamber was repeatedly flipped and imaged, as described in Sec.\ \ref{methods}. (b) Sample image from the side during a single sedimentation experiment of particles with $\chi=0$ and $\kappa=3$ (St+St+St). (c) Sample image during sedimentation of particles with $\chi=0.18$ and $\kappa=3$ (St+Al+Al).}
    \label{fig:3D experiments}
\end{figure}

\begin{figure}
    \centerline{\includegraphics[width = \columnwidth]{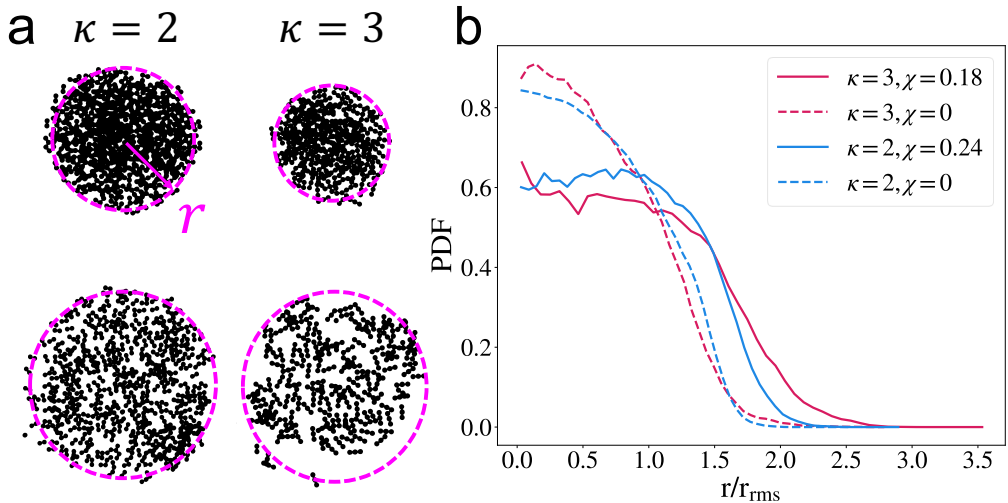}}
    \caption{(a) Sample images of particles resting on the bottom of the 3D chamber after sedimentation. The left column shows $\kappa=2$ particles with $\chi=0$ (St+St; top) and $\chi=0.24$ (St+Al, bottom).  The right column shows $\kappa=3$ particles with $\chi=0$ (St+St+St; top) and $\chi=0.18$ (St+Al+Al, bottom). We quantified the radial distribution of black pixels at a different radii $r$ from the center of the sedimented pattern, as indicated by the magenta arrow. (b) Radial probability density function (PDF) of black pixels from the images. The radius has been normalized by $r_{rms}$ for $\chi$ = 0 (Eq.\ \ref{rms}). The legend indicates the values of $\kappa$ and $\chi$ for the different PDFs. Each curve was produced using data from 50 post-sedimented images.}
    \label{fig:3D Data}
\end{figure}


Although the effective repulsion between our prolate particles is apparent in a confined, quasi-2D environment, it is possible that the dynamical evolution of these particles in 3D could hinder the repulsion since the particles have more motional degrees of freedom. Figure \ref{fig:3D experiments} illustrates the experimental procedure, described in Sec.\ \ref{methods}, where particles are sedimented repeatedly and their resulting spatial distribution is imaged after each repeated experiment. For $\chi=0$, particles tended to cluster during sedimentation, resulting in a rapid increase in their velocity due to mutual drag reduction at finite distances. For $\chi=0.18$, there is a visible alignment of particle to the direction of gravity (vertical), and a broader spatial distribution with less clustering. 

To quantify the post-sedimentation spatial distribution of particles, one would ideally extract the center of mass position of each particle and calculate the radial distribution function of their positions. However, after sedimentation we found that particles often overlapped by stacking in the vertical direction, making identifying the center of mass impossible. Instead, we choose to threshold the images so that particles became black pixels, and the background became white. Samples of these images are shown in Fig. \ref{fig:3D Data}. We then calculated the radial distribution function of the positions of the black pixels. This was done by first finding the center of mass of all black pixels, corresponding to $r=0$ in each image, and then binning pixel positions radially along $r$. We divided the number of pixels in each bin range by the area of the annulus associated with the bin. In order to compare between $\kappa=2$ and $\kappa=3$, we also normalized the radial positions by the root mean squared radius of the data for $\chi=0$, calculated by:
\begin{equation}
r_{rms} = \sqrt{\dfrac{1}{N} \sum_r r^2}.
\label{rms}
\end{equation}
Here, the sum runs over every black pixel in all 50 images associated with $\chi=0$, and $N$ is the sum total of all black pixels from all images with $\chi=0$. 

The resulting distribution functions are shown in Fig.\ \ref{fig:3D Data}. After normalizing the radial position, we see that all data for $\chi=0$ collapses onto the same distribution (dashed lines). As expected, when $\chi>0$ (solid lines), these distributions broaden due to the net repulsion between the particles. Furthermore, particles with larger $\kappa$ and smaller non-zero $\chi$ should experience a larger respulsion, as predicted by Eq.\ \ref{sepeq}. This is consistent with our data, since the distribution for $\kappa=3$, $\chi=0.18$ is broader than for $\kappa=2$, $\chi=0.24$. We note that because the initial state of each round of sedimentation was set by the final state of the previous one, the sequential images of the final sedimentation pattern were not statistically independent. Nevertheless, we don't expect these effects to qualitatively change our results, and taken together, Fig.\ \ref{fig:3D experiments} and Fig.\ \ref{fig:3D Data} confirm that the effective pairwise repulsion between mass polar particles also suppresses clumping in 3D.

\section{Conclusion}

Particles with mass polarity are forced to align with the direction of gravity during sedimentation. This alignment arises because the center of mass is displaced from the geometric center of each particle, resulting in a net torque imposed by the fluid flow. Our work examined the motion of single particles sedimenting in a viscous fluid, and we derived a simple analytic expression for the their motion in a quasi-2D environment. Fitting trajectories of the individual particles allowed us to reconstruct the parameters of the mobility matrix. When two or more prolate particles are present, we showed that they experience a mutual repulsion, as first described by \citet{Goldfriend2017}. Surprisingly, this repulsion is strongest for small values of $\chi$, i.e. when the center of mass is only slightly displaced from the center of geometry. The repulsion also increases as the particles become more asymmetric (more prolate, large $\kappa$). We also showed that this overall repulsion persists in 3D experiments with hundreds of particles.   

There are still many open questions facilitated by this work. First, \citet{Goldfriend2017} showed that there should be a mutual attraction between particles for $\kappa<1$. We found that our particle fabrication method, i.e. gluing individual spheres together, did not easily lend itself to making oblate particles with $\kappa<1$. Such particles would cluster rapidly during sedimentation, and may dramatically increase the overall sedimentation rate of a suspension of particles. Additionally, \citet{Goldfriend2017} predicted the existence of hyperuniformity in the density distribution of a sedimenting suspension. Our experimental results in 3D demonstrate a net repulsion and a more uniform concentration, yet we would need many more particles with accurate tracking in 3D to quantify hyperuniformity. One alternative route could be simulating many particles efficiently with a parameterized interaction based on our results. We hope our simplified mobility matrix may serve as a starting point for such idealized simulations of many interacting particles.

We sincerely thank Maciej Lisicki, Tom Witten, and Haim Diamant for stimulating discussions. This material is based upon work supported by the National Science Foundation under Grant No. 2025795. Declaration of Interests. The authors report no conflict of interest.

\bibliographystyle{jfm}
\bibliography{sedimentation}

\end{document}